\documentclass{aa}

\usepackage{graphicx}
\usepackage{mathrsfs}  
\usepackage{natbib}
\usepackage{amssymb}
\bibpunct{(}{)}{;}{a}{}{,}
\usepackage{caption}
\usepackage[varg]{txfonts}
\usepackage[section]{placeins}
\usepackage[english]{babel}
\usepackage{url}
\usepackage{arydshln}
\usepackage{color}

\def\phn{\phantom{0}}

\def\tablecomments#1{\par\smallskip\noindent Notes. #1}

\authorrunning{Magee et al.}
\titlerunning{PS1-12bwh}

\begin{document} 
            
    \title{Growing evidence that SNe Iax are not a one-parameter family: }
\subtitle{the case of PS1-12bwh}
		\author{M. R. Magee\inst{1}{\thanks{E-mail: mmagee37@qub.ac.uk}}
        \and R. Kotak\inst{1}
        \and S. A. Sim\inst{1}
        \and D. Wright\inst{1}
        \and S. J. Smartt\inst{1}
        \and E. Berger\inst{2}
        \and R. Chornock\inst{3}
        \and R. J. Foley\inst{4}
        \and D. A. Howell\inst{5,6}
        \and N. Kaiser\inst{7}
        \and E. A. Magnier\inst{7}
        \and R. Wainscoat\inst{7}
        \and C. Waters\inst{7}
            }
	\institute{Astrophysics Research Centre, School of Mathematics and Physics, Queen's University Belfast, Belfast, BT7 1NN, UK
    \and Harvard-Smithsonian Center for Astrophysics, 60 Garden Street, Cambridge, Massachusetts 02138, USA
     \and Astrophysical Institute, Department of Physics and Astronomy, 251B Clippinger Lab, Ohio University, Athens, OH 45701, USA   
     \and Department of Astronomy and Astrophysics, University of California, Santa Cruz, CA 95064, USA   
     \and Las Cumbres Observatory Global Telescope, 6740 Cortona Dr., Suite 102, Goleta, CA 93111, USA 
     \and Department of Physics, University of California, Santa Barbara, Broida Hall, Mail Code 9530, Santa Barbara, CA 93106-9530, USA
     \and 	Institute for Astronomy, University of Hawaii at Manoa, Honolulu, HI 96822, USA   
}
   \date{Received -	- -; accepted - - - }

 
  \abstract{In this study, we present observations of a type Iax supernova, PS1-12bwh, discovered during the Pan-STARRS1 3$\pi$-survey. Our analysis was driven by previously unseen pre-maximum, spectroscopic heterogeneity. While the light curve and post-maximum spectra of PS1-12bwh are virtually identical to those of the well-studied type Iax supernova, SN~2005hk, the $-$2\,day spectrum of PS1-12bwh does not resemble SN~2005hk at a comparable epoch; instead, we found it to match a spectrum of SN~2005hk taken over a week earlier ($-$12\,day). We are able to rule out the cause as being incorrect phasing, and argue that it is not consistent with orientation effects predicted by existing explosion simulations. To investigate the potential source of this difference, we performed radiative transfer modelling of both supernovae. We found that the pre-maximum spectrum of PS1-12bwh is well matched by a synthetic spectrum generated from a model with a lower density in the high velocity ($\gtrsim$6000 km~s$^{-1}$) ejecta than SN~2005hk. The observed differences between SN~2005hk and PS1-12bwh may therefore be attributed primarily to differences in the high velocity ejecta alone, while comparable densities for the lower velocity ejecta would explain the nearly identical post-maximum spectra. These two supernovae further highlight the diversity within the SNe Iax class, as well as the challenges in spectroscopically identifying and phasing these objects, especially at early epochs. 
}


   \maketitle
%

\section{Introduction}
\label{sect:intro}

Although there is diversity among SNe Ia, they are generally well described as a one parameter family. For most ``Branch-normal" SNe Ia \cite[][]{branch--ia--groups}, parameters such as the light curve shape are highly correlated with the peak absolute magnitude, and hence also with each other. Thus, the amount of $^{56}$Ni synthesized in the explosion is the primary driver of observational characteristics, and it is this property that underpins the use of SNe Ia as distance indicators \citep{branch-tammann--ia-dist}. The same cannot be said for the type Iax SNe, also known as `02cx-like' supernovae after its prototype, SN~2002cx \citep{02cx--orig}. In analogy to the SNe Ia, it has been suggested that SNe Iax are also the result of the thermonuclear disruption of a white dwarf. Indeed, it may be that pulsational delayed detonations underlie both SNe Ia \citep[e.g.][]{hoeflich-95, dessart-11fe} and SNe Iax \citep[SN~2012Z; ][]{comp--obs--12z}. However, unlike the SNe Ia, type Iax SNe do not exhibit tight correlations in their photometric properties, and are markedly different spectroscopically. This may suggest that SNe Iax result from a different explosion mechanism. In particular, it has been proposed that SNe~Iax might be associated with the pure deflagration of a white dwarf \citep{read--02cx--spectra,02cx--late--spec}. Within this scenario, particular attention has been paid to models that only partially disrupt the white dwarf \citep{kicked--remnants,3d--deflag--sim--rem}. 
\par
SNe Iax are distinct from normal SNe Ia in both their photometric and spectroscopic properties. Most notably, SNe Iax can be fainter than normal SNe Ia \citep{02cx--orig} by up to five magnitudes \citep{low--ccsn, obs--08ha}. They also display large variations in their light curve shapes, which show a  broader range of decline rates $0.3 \lesssim \Delta$m$_{15}$(r) $\lesssim$ 1.1 \citep[SN~2009ku, SN~2008ha,][respectively]{obs--09ku,obs--10ae}, and rise times, from $\sim$10\,d \citep[SN~2010ae, ][]{obs--10ae, 15h} to $\sim$22\,d \citep[SN~2005hk,][]{05hk--deflag?, 15h}, when compared to normal SNe Ia. The fact that many SNe Iax have neither well-constrained peak magnitudes nor decline rates, and even fewer have secure rise time measurements, has further added to the difficulty in interpreting the observed behaviour of SNe Iax. The range of observational diversity exhibited by SNe Iax has not been completely characterized, while the range of expected behaviour remains to be fully explored, particularly in the context of SNe Iax explosion models.
\par
Some of the diversity observed in SNe Iax is illustrated by their maximum light spectra. Absorption features due to intermediate mass elements (IMEs, e.g. silicon, sulphur) are usually apparent, but with strengths that vary substantially from one supernova to another, and are never as strong as in normal SNe Ia. Pre-maximum spectra are unavailable for the vast majority of SNe Iax, but based on the maximum light spectra, one may reasonably expect even greater disparity. There is also some evidence that the observed properties of SNe Iax are not controlled by a single parameter. For example, SN~2009ku displayed exceptionally low velocities for its peak brightness \citep{obs--09ku}.
\par
Here we present a comparative analysis focussed primarily on the pre-maximum data of the type Iax supernovae PS1-12bwh and SN~2005hk. The study was motivated primarily by the striking differences in pre-maximum spectra accompanied by nearly identical photometric evolution. In what follows, we use the term ``ejecta'' to refer specifically to unbound material, unless stated otherwise.

\begin{figure}[!t]
\centering
\includegraphics[width=\columnwidth]{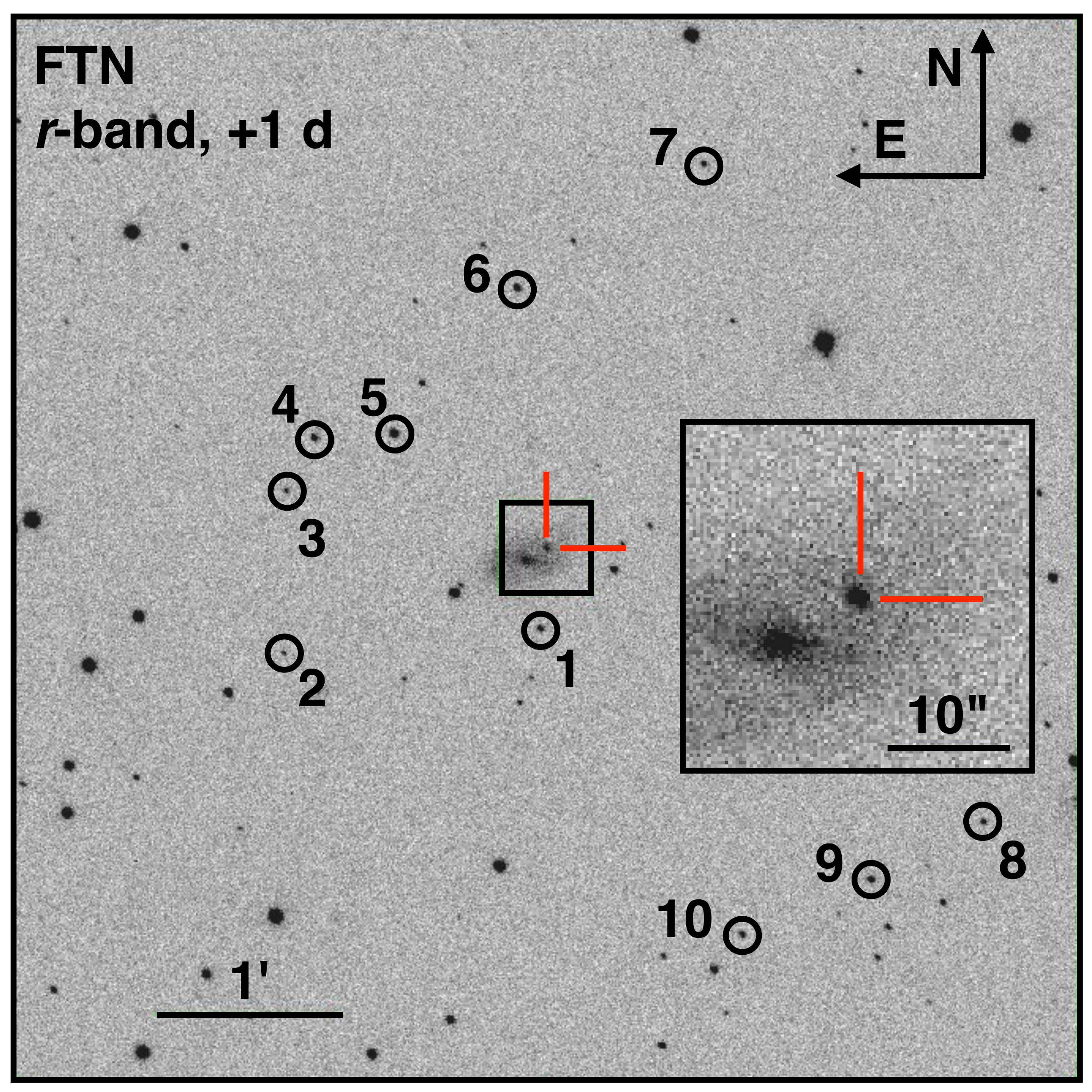}
\caption{$r-$band image of PS1-12bwh taken with Faulkes Telescope North (FTN) approximately one day post $r-$band maximum. PS1-12bwh (marked by red dashes) is located at $\alpha = 07^\mathrm{h}09^\mathrm{m}24^\mathrm{s}.29, \, \delta = +39\degr06'15.8\farcs$, and lies approximately 4$''$N 6$''$W from the nucleus of CGCG 205-201.
Sequence stars used to calibrate photometry are marked and listed in Table~\ref{tab:12bwh---std}. 
}
\label{fig:12bwh--field}
\centering
\end{figure}

%

%

\section{Observations \& Data Reduction}
\label{sect:obsandreduc}
PS1-12bwh was discovered during routine operations of the 3$\pi$ all sky survey by the Panoramic Survey Telescope and Rapid Response System \citep[hereafter PS1, ][]{ps1--kaiser, magnier--2013} on 2012 Oct. 17. During  the early phases of PS1 operations, we ran the Faint Galaxy Supernova survey in which we cross matched sources found in the nightly (undifferenced) images  with faint galaxies identified in the Sloan Digital Sky Survey. The first transients were reported by \cite{valenti--2010} and the survey procedures are detailed by \cite{inserra--2013}.  PS1-12bwh was discovered in this way, and reported by \cite{12bwh--atel}.
\par
\begin{figure}[!t]
\centering
\includegraphics[width=\columnwidth]{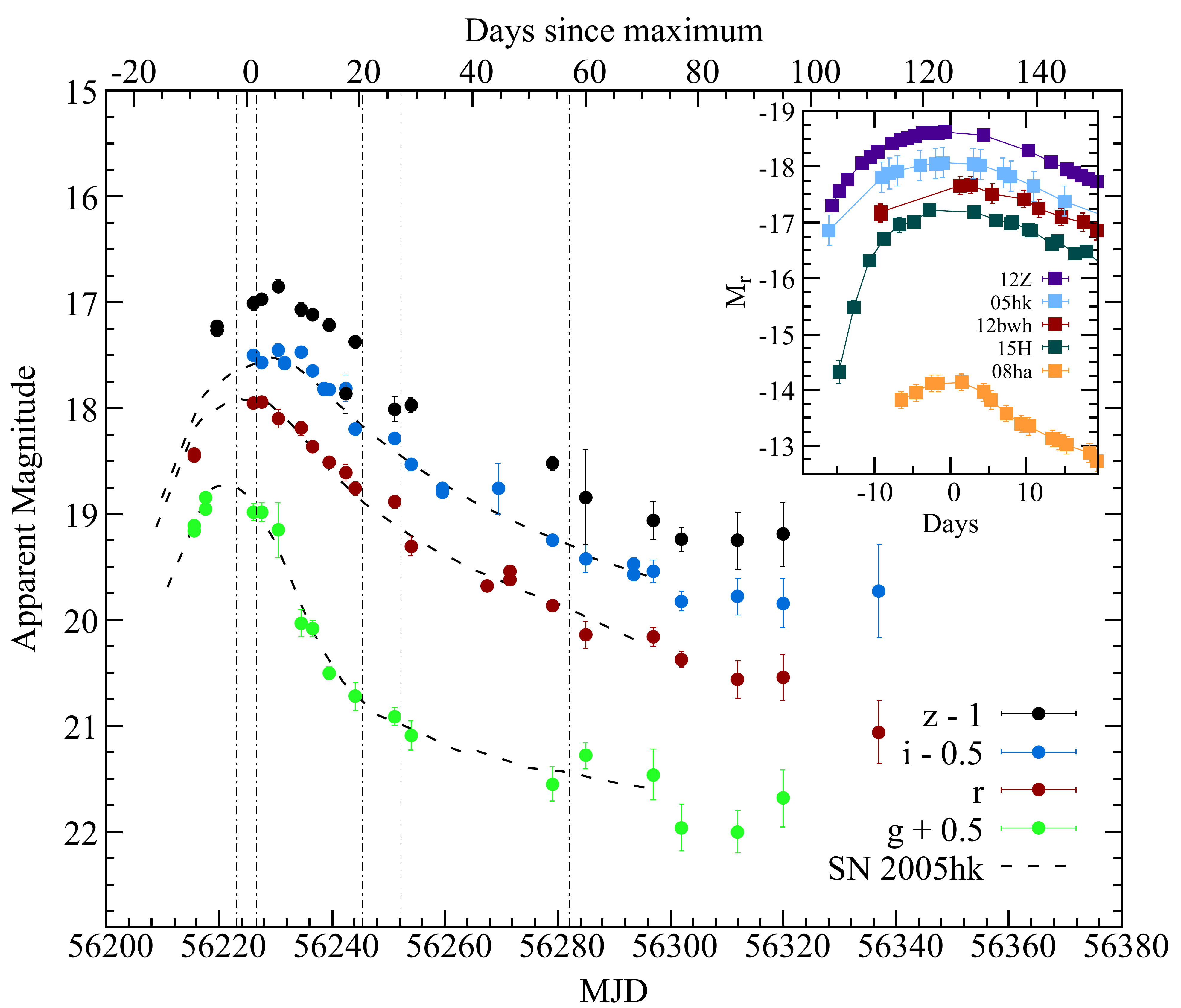}
\caption{Light curves of PS1-12bwh. Epochs of spectroscopic observations are marked with dot-dashed vertical lines. Light curves of SN~2005hk in the natural system \citep{comp--obs--12z} have been shifted in time and magnitude to match the light curves of PS1-12bwh, and are shown for comparison as dashed lines. Inset: Absolute $r-$band magnitude light curves of PS1-12bwh and comparison objects used throughout this paper. We limit this comparison to $\pm$20\,d relative to $r-$band maximum as an indication of the rise and decline of these objects. With the exception of SN~2008ha, all objects have comparable decline rates in the $r-$band, from 0.60 (PS1-12bwh) to 0.70 \citep[SN~2005hk, ][]{comp--obs--12z} and a spread of absolute peak magnitudes, from $M_r\sim-$17.3 \citep[SN~2015H, ][]{15h} to $M_r\sim -18.6$ \cite[SN~2012Z,][]{comp--obs--12z}. These values can be compared with $M_r \sim -19.24$ for SNe Ia \citep{cfa--3}.
Data sources are listed in Table~\ref{tab:objs}. 
}
\label{fig:12bwh--lc}
\end{figure}

Fig.~\ref{fig:12bwh--field} shows the field around PS1-12bwh. We supplemented the PS1 observations with targeted observations using the 2.0-m Liverpool Telescope \citep[LT, ][]{lt} and the 2.0-m Faulkes Telescope North \citep[FTN, ][]{lcogt}. Our complete set of photometric observations is shown in Table~\ref{tab:12bwh--phot} and Fig.~\ref{fig:12bwh--lc}. 
\par
PS1, FTN, and LT each have custom-built data reduction pipelines that process the images, and apply basic calibrations automatically \citep[][respectively]{ipp--2, schlafly--2012, lcogt, lt}. Given that PS1-12bwh occurred at an offset of only $\sim$3\,kpc (projected distance) from the nucleus of its host galaxy, CGCG 205-201, we opted to perform the photometry measurements using template-subtracted difference images. The images were first aligned using point sources in common between the images; the template image was then subtracted using HOTPANTS\footnote{http://www.astro.washington.edu/users/becker/v2.0/hotpants.html}. For PS1 and FTN images, we used archival pre-explosion SDSS\footnote{www.sdss.org} images as templates for all bands. When PS1-12bwh had faded well below our detection limits ($\gtrsim$860\,d post-maximum), we acquired templates for the LT images. 

\begin{figure}[!t]
\centering
\includegraphics[width=\columnwidth]{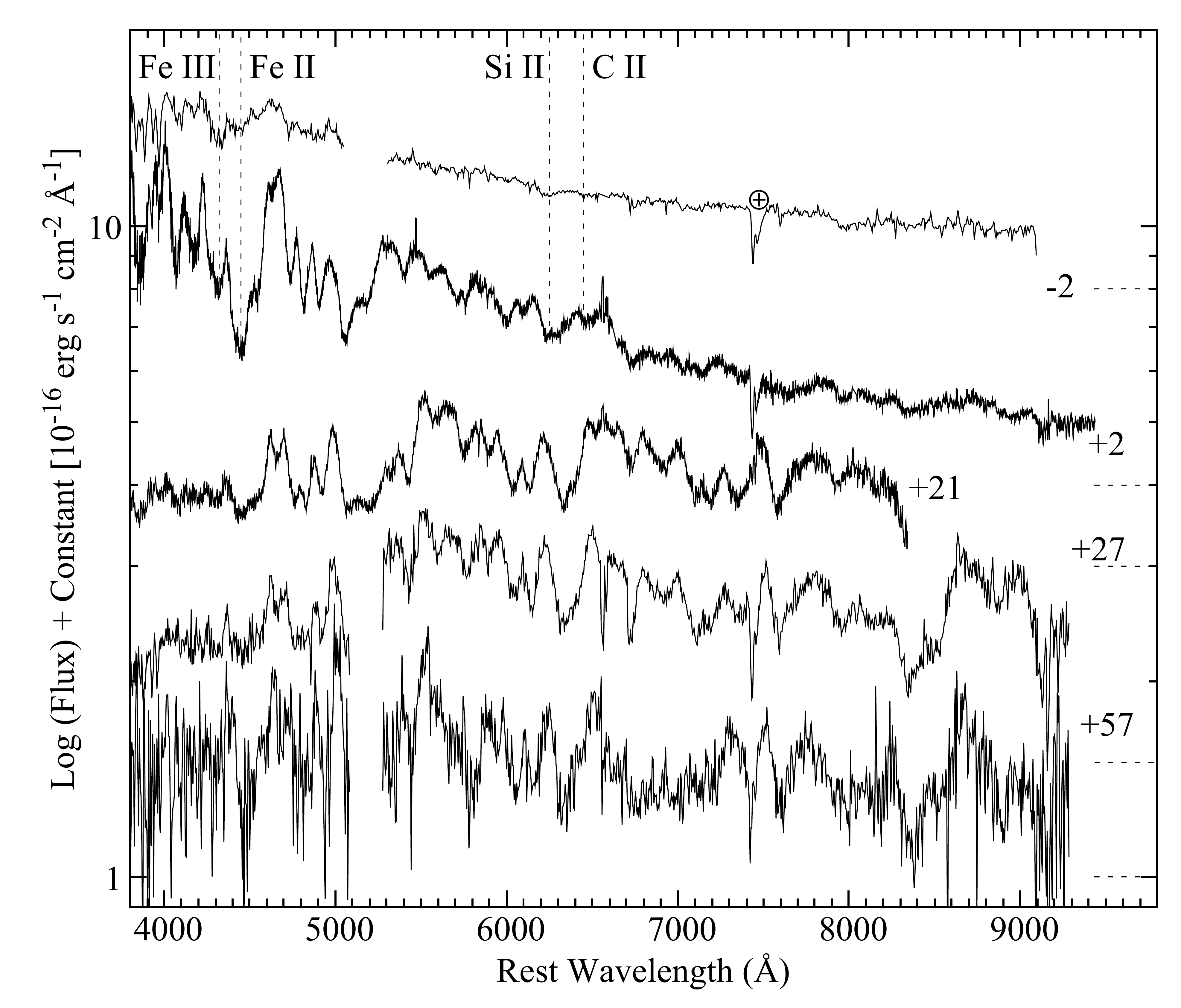}
\caption{Spectroscopic sequence of PS1-12bwh. All spectra have been corrected for redshift and galactic extinction. Epochs are given in days relative to an $r-$band maximum of MJD = 56224.9. For clarity, spectra have been offset vertically with the zero point of each offset marked by a dashed line. Features due to telluric absorption, \ion{Fe}{iii} $\lambda$4404, \ion{Fe}{ii} $\lambda$4555, \ion{Si}{ii} $\lambda$6355, and \ion{C}{ii} $\lambda$6580 are marked and narrow host emission lines have been removed. Gaps in the $-2$, $+27$, and $+57$\,d spectra are due to the instrument configuration (Table \ref{tab:12bwh--specseq}).
}
\label{fig:12bwh--spec-log}
\end{figure}

Photometry on all images was performed using SNOoPY\footnote{http://sngroup.oapd.inaf.it/snoopy.html}, a custom built IRAF\footnote{The Image Reduction and Analysis Facility (IRAF) is maintained and distributed by the Association of Universities for Research in Astronomy, under the cooperative agreement with the National Science Foundation.} package designed for point spread function (PSF) fitting photometry. SNOoPY uses a selection of stars in the field to build up a reference PSF; these stars are listed in Table~\ref{tab:12bwh---std} and shown in Fig.~\ref{fig:12bwh--field}. The SN is extracted from the image using this reference PSF and an initial background estimate. A new estimate of the background is then derived using the SN-subtracted image. This process is repeated until the residuals are minimized, yielding the final SN magnitude. Instrumental magnitudes of the standard stars are then calibrated to the SDSS magnitudes yielding zero points and colour terms which are used to calculate the apparent magnitude of the SN. Using the procedure described by \cite{ps1--phot--sys}, we converted our observed magnitudes in the PS1 filters to the standard SDSS system.

\par
Approximately five days after discovery, a spectrum obtained with the ISIS spectrograph mounted on the William Herschel Telescope (WHT) showed PS1-12bwh to be a SN Iax around maximum light, with notable similarities to SN~2008ae \citep{12bwh--atel}. We acquired two additional spectra, also from WHT + ISIS, roughly one and two months following the initial classification. All WHT spectra were obtained using the same instrumental set-up: 5300\,{\AA} dichroic with the R300B and R158R gratings.  We additionally obtained two spectra; one with the GMOS spectrograph mounted on the Gemini North telescope, and one from the Multiple Mirror Telescope (MMT) with the 300GPM grating. A log of the spectroscopic observations is presented in Table~\ref{tab:12bwh--specseq}, and our full spectroscopic sequence is shown in Fig.~\ref{fig:12bwh--spec-log}.

\par
Spectroscopic data were reduced using standard IRAF routines. In addition to the flux calibration using spectroscopic flux standards, we adjusted the flux levels of our spectra to the photometry. We did this by calculating synthetic magnitudes of our spectra using SMS \citep[Synthetic Magnitudes from Spectra, ][]{inserra-2016} and scaled these such that the synthetic magnitudes match the photometric measurements. 
\par
Using narrow emission features due to the host galaxy present in our earliest spectrum, we derive a redshift of $z = 0.0228\pm0.0005$ to PS1-12bwh, corresponding to a distance modulus of $\mu$=34.91$\pm$0.15 ($D_L$ = 96 Mpc), assuming H$_0$ = 70.0 km s$^{-1}$ Mpc$^{-1}$, $\Omega_M$ = 0.3, and $\Omega_{\Lambda}$ = 0.7. These values are consistent with the NED values for CGCG 205-201.

\section{Analysis}
\label{sect:analysis}

\subsection{Reddening}
\label{sect:red}
Unlike normal SNe Ia, SNe Iax as a group do not follow a well defined colour evolution \citep{foley--13}. This presents a challenge in determining the level of  extinction due to the host galaxy, as one must turn to alternative methods. Empirical relations between the strength of \ion{Na}{i} absorption features and extinction are commonly used to estimate the extinction towards SNe. The \ion{Na}{i} doublet is only visible in the $-$2\,d spectrum. Prior to measuring the equivalent widths, we normalised the spectrum by fitting and dividing by the pseudo-continuum using a low-order polynomial. This resulted in values of D1 =  0.42$\pm$0.10\,\AA ~and D2 = 0.56$\pm$0.07\,\AA. Using the equations of \cite{na1d-red-2}, this provides an estimate for the host extinction of PS1-12bwh of $E(B-V)_{\mathrm{host}} = 0.20\pm$0.07 mag. Combining this with the extinction in the direction of PS1-12bwh provided by NED, this results in a total extinction value of $E(B-V)_{\mathrm{total}} = 0.26\pm$0.07 mag, assuming $R_V$ = 3.1, which we adopt throughout this study.

\subsection{Photometry}
\label{sect:photometric}

In Fig.~\ref{fig:12bwh--lc}, we present the full light curve of PS1-12bwh. It includes observations before or around maximum light in each filter, and extends to over 100\,d post $r$--band maximum. In keeping with observations of other SNe Iax, PS1-12bwh does not show any sign of a secondary maximum in the $i$ or $z$ bands. 
\par
By fitting low-order polynomials, we derived the light curve parameters for PS1-12bwh in each of the $griz$ filters. We find that it attained its peak $r$--band value of +17.93$\pm$0.04 on MJD = 56224.9$\pm$1.3. This corresponds to a peak absolute $r$-band magnitude of $-$17.69$\pm$0.24 using our derived distance modulus and extinction. Decline rates in each filter are calculated from the peak magnitude and the corresponding magnitude 15\,d later; the values are listed in Table~\ref{tab:12bwh--lc--param}. The uncertainty in the absolute magnitude of PS1-12bwh is dominated by the uncertainty in host extinction. From here onwards, all references to the epoch of maximum light refer to the date of $r$--band maximum. 

\par
As can be seen from Table~\ref{tab:12bwh--lc--param} and Fig.~\ref{fig:12bwh--lc}, PS1-12bwh displays a remarkably similar photometric evolution to SN~2005hk. As a check, we also derived light curve parameters for PS1-12bwh using SN~2005hk as a template (Fig.~\ref{fig:12bwh--lc}), and find them to be consistent with those reported in Table~\ref{tab:12bwh--lc--param}. Unfortunately, there are no observations of the field of PS1-12bwh in any filter within 100\,d prior to discovery, so we are therefore unable to set a robust limit on its explosion epoch. 
%
%

\begin{table*}
\centering
\caption{Photometric journal for  PS1-12bwh}\tabularnewline
\label{tab:12bwh--phot}\tabularnewline
\begin{tabular}{llllllll}
\hline\hline
\tabularnewline[-0.25cm]
Date & MJD & Phase & $g$ & $r$ & $i$ & $z$ & Telescope  \tabularnewline
 &  & (days) & (mag) & (mag) & (mag) & (mag) & (instrument) \tabularnewline
\hline
\hline
\tabularnewline[-0.25cm]
2012 Oct. 15 &	56215.60 & \phn\phn$-$9		& 18.61(0.04) 	& 18.45(0.02) 	& ... 			& ... 			& PS1 (GPC1) 	\tabularnewline
			&	         & 				& 18.66(0.04) 	& 18.43(0.02) 	& ... 			& ... 			& PS1 (GPC1) 	\tabularnewline
2012 Oct. 17 &	56217.58 & \phn\phn$-$7		& 18.34(0.03) 	& ... 			& ... 			& ... 			& PS1 (GPC1) 	\tabularnewline
			&	         & 				& 18.45(0.03) 	& ... 			& ... 			& ... 			& PS1 (GPC1) 	\tabularnewline
2012 Oct. 19 &	56219.62 & \phn\phn$-$5		& ... 			& ... 			& ... 			& 18.22(0.03) 	& PS1 (GPC1) 	\tabularnewline
			&	         & 				& ... 			& ... 			& ... 			& 18.26(0.03) 	& PS1 (GPC1) 	\tabularnewline
2012 Oct. 26 &	56226.15 & \phn\phn+1	& 18.48(0.08) 	& 17.95(0.05) 	& 18.00(0.04) 	& 18.01(0.07) 	& LT (RATCam) 	\tabularnewline
2012 Oct. 27 &	56227.55 & \phn\phn+3 	& 18.48(0.09) 	& 17.94(0.04) 	& 18.07(0.04) 	& 17.97(0.04) 	& FTN (FS)	 	\tabularnewline
2012 Oct. 30 &	56230.41 & \phn\phn+6	& 18.65(0.26) 	& 18.10(0.09) 	& 17.95(0.06) 	& 17.85(0.07) 	& FTN (FS)	 	\tabularnewline
2012 Oct. 31 &	56231.55 & \phn\phn+7	& ... 			& ... 			& 18.08(0.02) 	& ... 			& PS1(GPC1) 	\tabularnewline
			&	         & 				& ... 			& ... 			& 18.07(0.02) 	& ... 			& PS1(GPC1) 	\tabularnewline
2012 Nov. 03 &	56234.56 & \phn+10		& 19.53(0.13)	& 18.19(0.06) 	& 17.97(0.05) 	& 18.07(0.07) 	& FTN (FS)	 	\tabularnewline
2012 Nov. 05 &	56236.55 & \phn+12		& 19.58(0.08) 	& 18.36(0.04) 	& 18.15(0.04) 	& 18.12(0.05) 	& FTN (FS)	 	\tabularnewline
2012 Nov. 07 &	56238.51 & \phn+14		& ... 			& ... 			& 18.31(0.02) 	& ... 			& PS1(GPC1)	 	\tabularnewline
			&	         & \phn			& ... 			& ... 			& 18.32(0.02) 	& ... 			& PS1(GPC1) 	\tabularnewline
2012 Nov. 08 &	56239.53 & \phn+15		& 20.00(0.06) 	& 18.51(0.03) 	& 18.32(0.03) 	& 18.21(0.05) 	& FTN (FS)	 	\tabularnewline
2012 Nov. 11 &	56242.40 & \phn+18		& ... 			& 18.61(0.08) 	& 18.31(0.12) 	& 18.86(0.19) 	& FTN (FS) 		\tabularnewline
2012 Nov. 13 &	56244.18 & \phn+19		& 20.22(0.13) 	& 18.76(0.06) 	& 18.70(0.05) 	& 18.37(0.06) 	& LT (RATCam) 	\tabularnewline
2012 Nov. 20 &	56251.14 & \phn+26		& 20.41(0.08) 	& 18.88(0.06) 	& 18.78(0.06) 	& 19.01(0.12) 	& LT (RATCam) 	\tabularnewline
2012 Nov. 23 &	56254.00 & \phn+29		& 20.59(0.14) 	& 19.30(0.09) 	& 19.03(0.06) 	& 18.97(0.07) 	& LT (RATCam) 	\tabularnewline
2012 Nov. 28 &   56259.51 & \phn+35		& ... 			& ... 			& 19.26(0.03) 	& ... 			& PS1(GPC1) 	\tabularnewline
			&	     	 & \phn			& ... 			& ... 			& 19.29(0.04) 	& ... 			& PS1(GPC1) 	\tabularnewline
2012 Dec. 06 &	56267.56 & \phn+43		& ... 			& 19.68(0.04) 	& ... 			& ... 			& PS1(GPC1) 	\tabularnewline
2012 Dec. 08 &	56269.47 & \phn+45		& ... 			& ... 			& 19.26(0.24) 	& ... 			& FTN (EM01)	\tabularnewline
2012 Dec. 10 &	56271.54 & \phn+47		& ... 			& 19.62(0.04) 	& ... 			& ... 			& PS1(GPC1) 	\tabularnewline
			&	         & \phn			& ... 			& 19.54(0.04) 	& ... 			& ... 			& PS1(GPC1) 	\tabularnewline
2012 Dec. 18 &	56279.02 & \phn+54		& 21.05(0.16) 	& 19.86(0.05) 	& 19.75(0.04) 	& 19.52(0.07) 	& LT (RATCam) 	\tabularnewline
2012 Dec. 23 &	56284.94 & \phn+60		& 20.78(0.12) 	& 20.14(0.13) 	& 19.92(0.13) 	& 19.84(0.45) 	& LT (RATCam) 	\tabularnewline
2013 Jan. 01 &	56293.49 & \phn+69		& ... 			& ... 			& 20.07(0.06) 	& ... 			& PS1(GPC1) 	\tabularnewline
			&	         & \phn			& ... 			& ... 			& 19.97(0.06) 	& ... 			& PS1(GPC1) 	\tabularnewline
2013 Jan. 04 &	56296.98 & \phn+72		& 20.96(0.24) 	& 20.16(0.09) 	& 20.04(0.11) 	& 20.06(0.18) 	& LT (RATCam) 	\tabularnewline
2013 Jan. 09 &	56301.98 & \phn+77		& 21.46(0.22) 	& 20.37(0.07) 	& 20.32(0.09) 	& 20.24(0.11) 	& LT (RATCam) 	\tabularnewline
2013 Jan. 19 &	56311.95 & \phn+87		& 21.50(0.20) 	& 20.56(0.18) 	& 20.28(0.17) 	& 20.25(0.27) 	& LT (RATCam) 	\tabularnewline
2013 Jan. 27 &	56319.93 & \phn+95		& 21.18(0.27) 	& 20.54(0.22) 	& 20.34(0.23) 	& 20.19(0.30) 	& LT (RATCam) 	\tabularnewline
2013 Feb. 13 &	56336.96 & +112			& ... 			& 21.06(0.30) 	& 20.23(0.44) 	& ...		 	& LT (RATCam) 	\tabularnewline
\hline
\end{tabular}
\begin{flushleft}
\tablecomments{Phases are given relative to the estimated $r$-band maximum of MJD = 56224.9 i.e., 6.6\,d later than $B$-band maximum \citep{05hk--deflag?}. }
\end{flushleft}
\end{table*}

\begin{table*}
\centering
\caption{Light curve parameters for PS1-12bwh compared to SN~2005hk}\tabularnewline
\label{tab:12bwh--lc--param}\tabularnewline
\begin{tabular}{llllll}\hline
\hline
\tabularnewline[-0.25cm]
SN & Filter  & Time of & Apparent peak & Absolute peak & Decline rate \tabularnewline
 &   & peak (MJD) &  (mag) & (mag) & $\Delta m_{15}$\,\, (mag) \tabularnewline
 
\hline
\hline
\tabularnewline[-0.2cm]
PS1-12bwh	&	$g$ & 56221.2(0.7) 	&	+18.24(0.06)	&	$-$17.65(0.30)	&	1.35(0.09)	\tabularnewline 
SN~2005hk	&	$g$ & 53685.4(0.1) 	&	+15.78(0.01)	&	$-$18.08(0.25)	&	1.36(0.01)	\tabularnewline 
\hline
PS1-12bwh	&	$r$ & 56224.9(1.3)	&	+17.93(0.04)	&	$-$17.69(0.24)	&	0.60(0.05)	\tabularnewline
SN~2005hk	&	$r$ & 53691.2(0.2) 	&	+15.68(0.01)	&	$-$18.07(0.25)	&	0.70(0.02)	\tabularnewline 
\hline
PS1-12bwh	&	$i$ & 56229.3(1.5)	&	+18.04(0.02)	&	$-$17.41(0.21)	&	0.67(0.04)	\tabularnewline
SN~2005hk	&	$i$ & 53695.3(0.7) 	&	+15.80(0.01)	&	$-$17.88(0.25)	&	0.60(0.01)		\tabularnewline 
\hline
\end{tabular}
\tablefoot{The light curve parameters of SN~2005hk are taken from \cite{comp--obs--12z}. }
\end{table*}

\par

\begin{figure}[!t]
\centering
\includegraphics[width=\columnwidth]{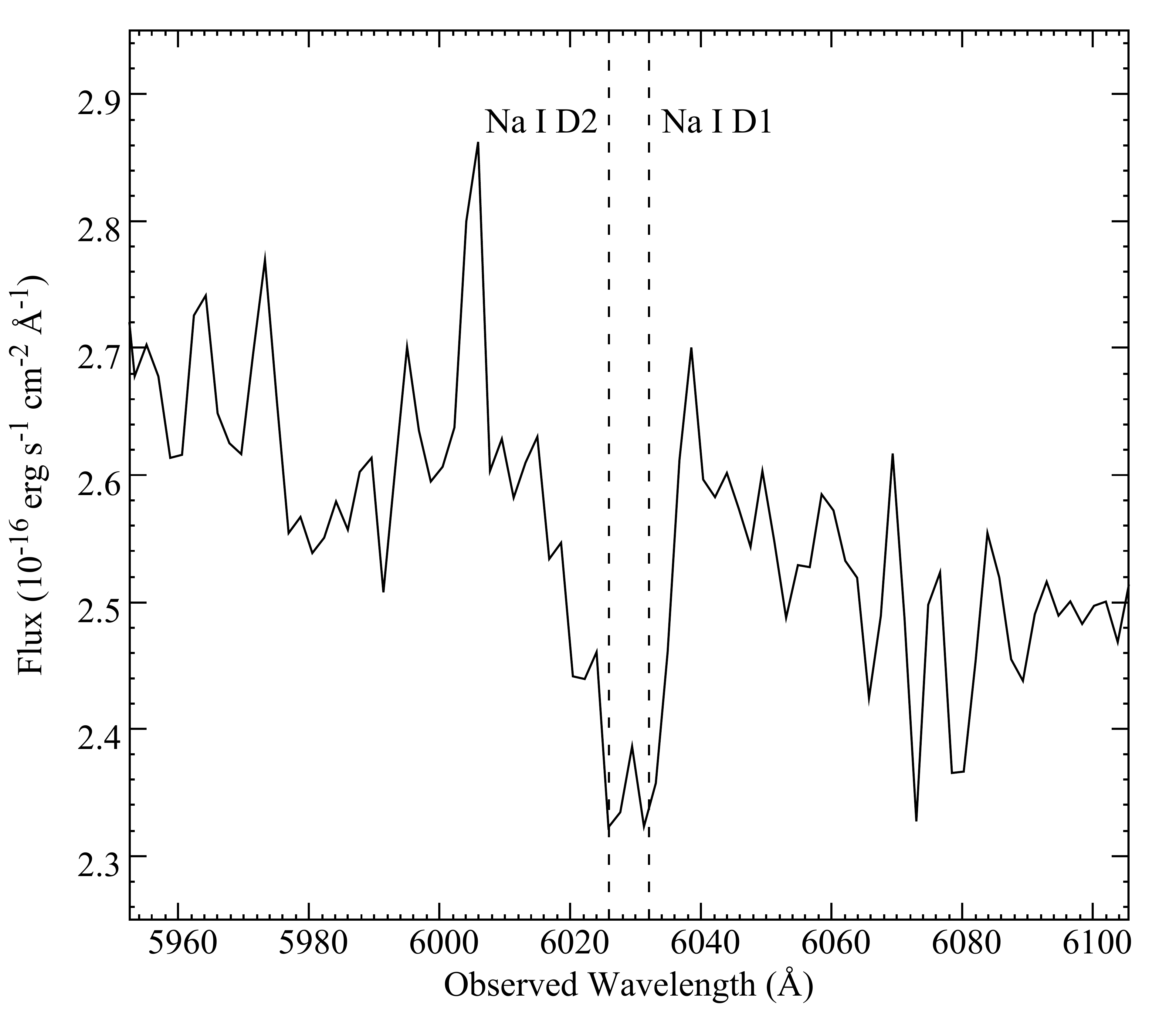}
\caption{Zoom-in of the $-$2d spectrum of PS1-12bwh. \ion{Na}{i} D absorption features are marked with vertical dashed lines in the frame of the host galaxy. }
\label{fig:12bwh--red}
\end{figure}


\subsection{Spectroscopy}
\label{sect:spectroscopic}

The full spectroscopic sequence for PS1-12bwh is shown in Fig.~\ref{fig:12bwh--spec-log}. 
\par

\subsubsection{Pre-maximum spectrum}
In spite of the virtually identical light curve evolution of PS1-12bwh and SN~2005hk, it can be seen from Fig.~\ref{fig:12bwh--residual} that their pre-maximum spectra are markedly different. Indeed, it is clear that the $-$2\,d spectrum of PS1-12bwh is more similar spectroscopically to the $-$12\,d SN~2005hk spectrum, or indeed similar spectra at even earlier epochs, than to the one at $-$3\,d. 
 
\par
At pre-maximum epochs, the spectra of SNe Iax generally show blue continua and few notable features, with exceptions being absorptions due to \ion{Fe}{iii} $\lambda$4404 and \ion{Fe}{ii} $\lambda$4555. Weak \ion{Si}{ii} $\lambda$6355 is also seen, but this feature is always far less prominent than in normal SNe Ia. However, features due to other IMEs are sometimes also apparent in SNe Iax \citep[e.g. \ion{S}{ii}, \ion{O}{i}, and \ion{Mg}{ii};][]{05hk--deflag?, early--late--08ha--obs, obs--2011ay}. Features due to \ion{Fe}{iii} $\lambda$4404, \ion{Fe}{ii} $\lambda$4555, and \ion{Si}{ii} $\lambda$6355 are observed in the $-$2\,d spectrum of PS1-12bwh and the $-$12\,d spectrum of SN~2005hk, although the \ion{Si}{ii} feature is somewhat stronger in SN~2005hk. From the \ion{Si}{ii} $\lambda$6355 feature, we infer velocities of $\sim$5700 km~s$^{-1}$, $\sim$6100 km~s$^{-1}$, and $\sim$5800 km~s$^{-1}$, respectively, for PS1-12bwh at $-$2\,d and SN~2005hk at $-12$ and $-3$\,d. Typical uncertainties in these measurements are $\pm$500 km~s$^{-1}$. The values for SN~2005hk reported above are consistent with those reported by \cite{05hk--deflag?}. We also tentatively identify absorption due to \ion{C}{ii} $\lambda$6580; this feature has been seen in the pre-maximum spectra of other SNe Iax \citep[e.g. ][]{foley--13}, and is indicative of unburned material in the ejecta. 
\par
As seen in Fig.~\ref{fig:12bwh--residual}, by $-$3\,d, many of the features observed in SN~2005hk have significantly increased in strength compared to the spectrum taken a week earlier, and bear little resemblance to the $-$2\,d spectrum of PS1-12bwh. This is particularly true of the \ion{Fe}{ii} $\lambda$4555, and \ion{Si}{ii} $\lambda$6355 features, which have dramatically increased in strength, although there is little change in the \ion{Fe}{iii} $\lambda$4404 absorption feature.

\subsubsection{Post-maximum spectra}
In Fig.~\ref{fig:12bwh--speccomp-max} we compare the $+2$\,d spectrum of PS1-12bwh to those of SNe 2005hk ($-3$\,d) and 2012Z (+2\,d). As there are no spectra available of SN~2005hk at +2\,d, we use SN~2012Z as a proxy, given its similarity to SN~2005hk \citep{comp--obs--12z}. It can be seen that the similarity of PS1~12bwh to spectra of SN~2005hk taken at earlier epochs persists to this post-maximum phase. 
\par
Despite the differences in the early spectra, at later epochs, the spectra of PS1-12bwh are remarkably similar to the spectra of other SNe Iax at comparable phases, including SN~2005hk, and in particular to those with similar decline rates. Objects shown in Fig.~\ref{fig:12bwh--speccomp-late} have decline rates ranging from $\sim$0.60 (PS1-12bwh) to 0.70 \cite[SN~2005hk; ][]{comp--obs--12z} in the $r-$band. As is typical of post-maximum SNe Iax spectra, the weak signatures of intermediate mass elements visible during the pre-maximum phases of PS1-12bwh have disappeared, and the spectra are now dominated by iron group species.
\par
Velocities measured from Fe-group features show a steady decline from around maximum light to two months post-peak: from the \ion{Fe}{ii} $\lambda$6149 feature in the $+2$\,d spectrum, we find a velocity of $\sim$7500 km~s$^{-1}$.  Approximately one month post maximum light, the velocity has decreased to $\sim$4900 km~s$^{-1}$, and further to $\sim$4400 km~s$^{-1}$ by +57\,d. SN~2005hk shows a similar evolution in the velocity measured from this feature: $\sim$6300 km~s$^{-1}$ at +6\,d to $\sim$5000 km~s$^{-1}$ at $\sim$+30\,d, to $\sim$4000 km~s$^{-1}$ another month later. We note that at these epochs, features due to intermediate mass elements (such as \ion{Si}{ii} $\lambda$6355) are blended with features due to iron group elements, and are therefore not easily identifiable. Fig.~\ref{fig:12bwh--speccomp-late} shows a spectroscopic comparison of objects with similar $\Delta m_{15}(r)$ at post-maximum epochs. 

\subsubsection{Host galaxy metallicity}
The presence of narrow host galaxy features in our earliest spectrum allows us to estimate the metallicity of CGCG\,205-021. We do so by first fitting and subtracting the pseudo-continuum in the $-$2\,d spectrum. We then fit Gaussian profiles to the narrow \ion{H}{}$\alpha$ and [\ion{N}{ii}] $\lambda$6583 features observed. Using the empirical relation derived by \cite{pettini--metal} with the N2 index, we find a host metallicity of 12 + log(O/H) = 8.87$\pm$0.19 dex. 
Metallicity measurements for the host galaxies of other SNe Iax derived from the \citealt{pettini--metal} relation yield comparable values: 12 + log(O/H) = 8.16$\pm$0.15, 8.40$\pm$0.18, and 8.51$\pm$0.31 dex for SNe 2008ha, 2010ae, and 2012Z \citep[]{obs--08ha,obs--10ae,12z--oister}, respectively. SNe Iax therefore appear to show no preference for either sub- or super-solar\footnote{We assume a solar metallicity of 12 + log(O/H) = 8.69$\pm$0.05 \citep{solar--comp}.} environments. The existence of a link between the host galaxy metallicity and peak supernova magnitude could be used to shed light on the likely progenitor channels. Based on the metallicity estimates currently available in the literature, however, there does not appear to be a clear correlation.

%

 \section{Discussion}
 \label{sect:disc}
 
 \begin{figure*}[!t]
\centering
\includegraphics[width=\textwidth]{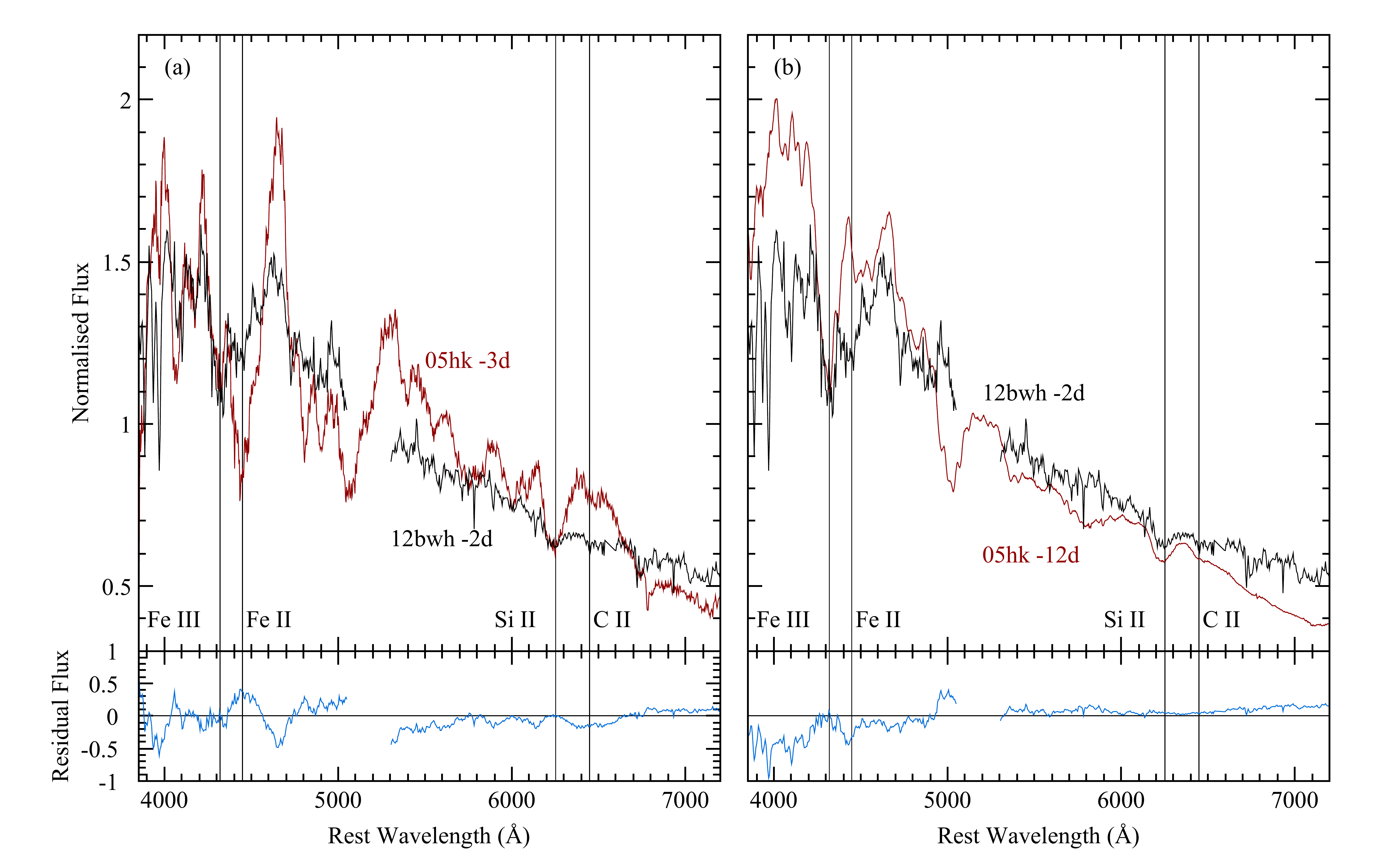}
\caption{Comparison of the $-2$\,d PS1-12bwh spectrum to two pre-maximum epochs of SN~2005hk. The spectrum of PS1-12bwh has been binned to $\Delta\lambda$ = 5~\AA~ and the narrow host emission lines have been removed. All spectra were normalised by the average continuum flux. The gap in the PS1-12bwh spectrum around 5200\,{\AA} is due to the dichroic used. The lower panels show the residuals resulting from taking the difference between the spectra in the respective upper panel, confirming that the $-2$\,d PS1-12bwh spectrum is a better spectroscopic match for the $-$12\,d SN~2005hk spectrum, than the similarly phased $-$3\,d spectrum. Features of interest are indicated by full vertical lines. The original data sources for spectra of SN~2005hk are given in Table~\ref{tab:objs}. 
}
\label{fig:12bwh--residual}
\end{figure*}

\begin{figure}[!t]
\centering
\includegraphics[width=\columnwidth]{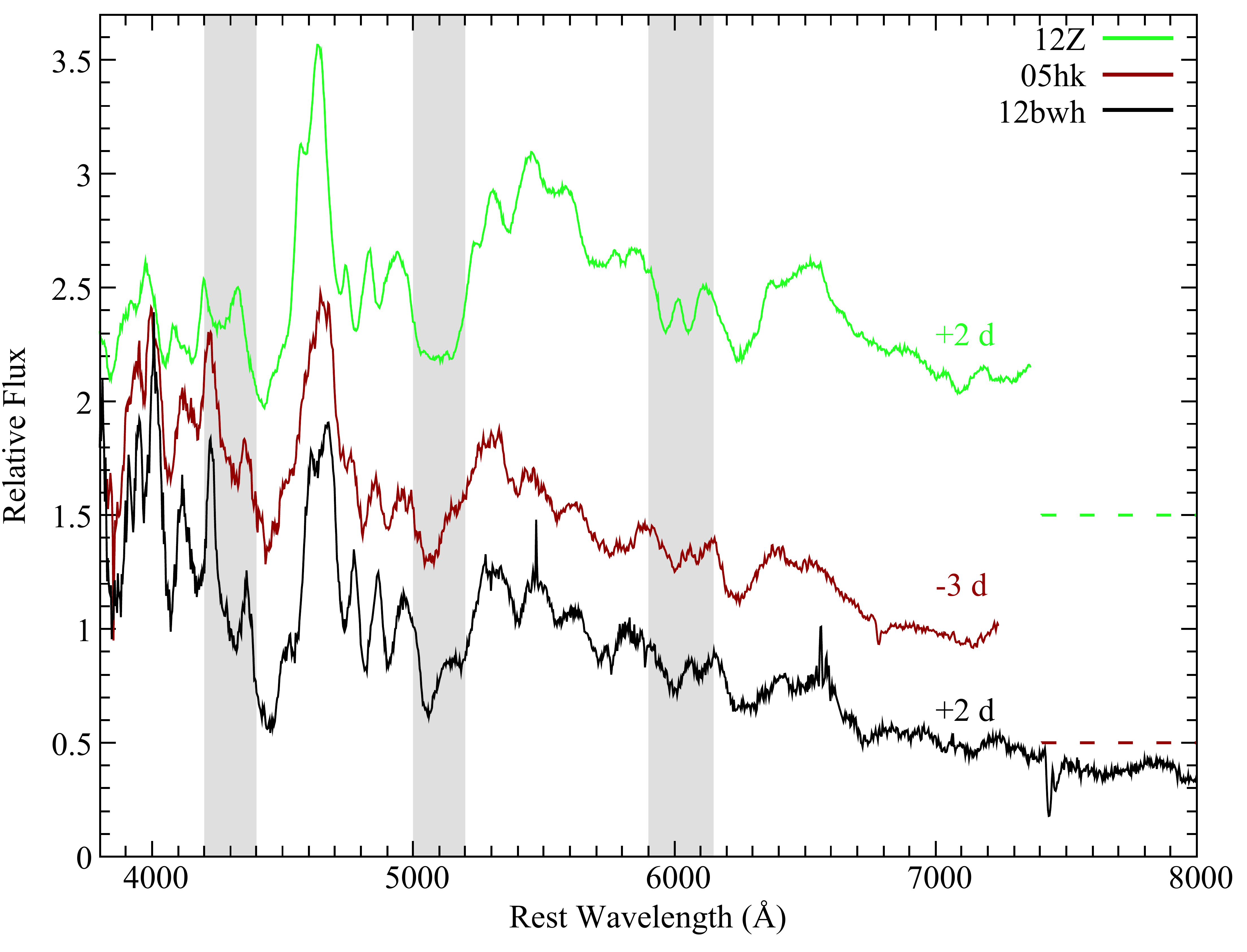}
\caption{Maximum light spectrum of PS1-12bwh compared to SN~2012Z at the same epoch and to SN~2005hk a day earlier. All spectra have been corrected for redshift and extinction, and have been rebinned to the same resolution ($\Delta\lambda = 3$\,\AA). SNe~2005hk and 2012Z have been offset vertically from PS1-12bwh for clarity, with the zero level of each offset marked by dashed lines.
Phases are given relative to the $r-$band maximum. The shaded areas highlight wavelength regions   
containing features with comparable strengths and shapes in the spectra of PS1-12bwh and SN~2005hk, but not SN~2012Z. PS1-12bwh at $+2$\,d appears to be qualitatively more similar to the pre-maximum SN~2005hk spectrum, than to the SN~2012Z spectrum at the same epoch. Data sources are listed in Table~\ref{tab:objs}.}
\label{fig:12bwh--speccomp-max}
\end{figure}

\begin{figure}[!t]
\centering
\includegraphics[width=\columnwidth]{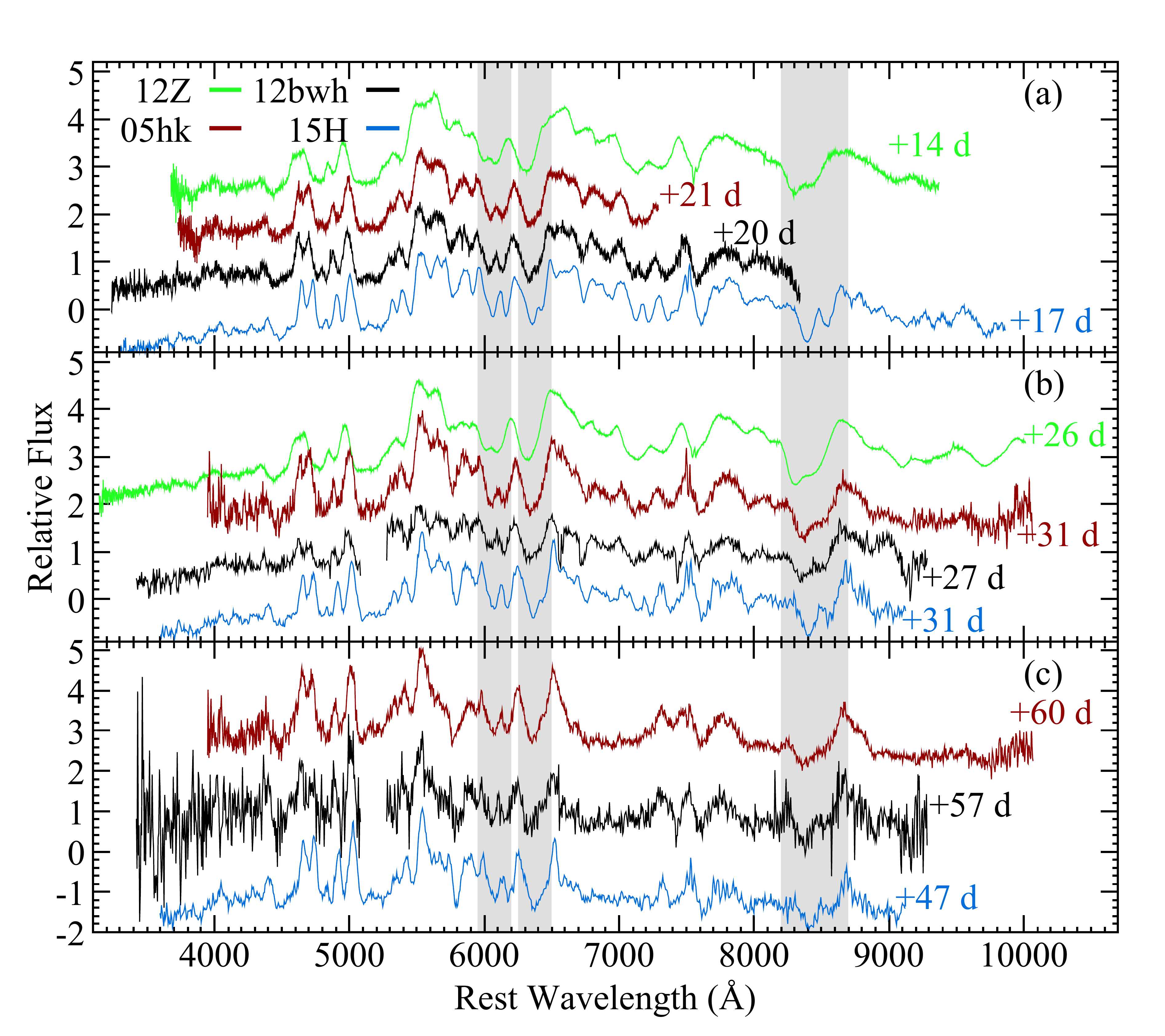}
\caption{Post-maximum spectra of PS1-12bwh compared with other SNe Iax at comparable epochs. All spectra have been corrected for redshift and reddening. Phases for PS1-12bwh are given relative to the $r$--band maximum (MJD = 56226.96). The WHT PS1-12bwh spectra have been binned to $\Delta\lambda$ = 5~\AA~and the narrow host emission lines have been removed. Shaded regions are discussed further in the main text. The data sources for all objects are listed in table~\ref{tab:objs}. The spectra  have been offset vertically from PS1-12bwh for clarity. All objects appear spectroscopically similar several weeks past maximum brightness, despite differences in velocity and peak brightness, from $M_r\sim-$17.3 \citep[SN~2015H, ][]{15h} to $M_r\sim -18.6$ \cite[SN~2012Z,][]{comp--obs--12z}. 
}
\label{fig:12bwh--speccomp-late}
\end{figure}

Here we investigate potential causes for the disparity between the $-$3\,d spectrum of SN~2005hk and the $-$2\,d spectrum of PS1-12bwh. 
\subsection{Epoch misidentification}
The first and most obvious potential cause is an error in the determination of the phase of either or both supernovae. The light curve of PS1-12bwh is shown in Fig.~\ref{fig:12bwh--lc}, with epochs of spectroscopic observations marked by dashed lines. As described in \S\ref{sect:photometric}, we estimated the $r$--band maximum to have occurred on MJD = 56224.9$\pm$1.3,  while our first spectrum was observed on MJD = 56223.1, i.e. only 1.8$\pm$1.3 days before $r$--band maximum. While there is inevitably some uncertainty in this phase, the $r$--band light curve of PS1-12bwh is clearly declining only a few days after our first spectrum was taken, and it is therefore not possible that this spectrum was taken close to 12\,d before maximum light. The $r$--band maximum of SN~2005hk is well constrained to MJD=53691.66$\pm$0.23 \cite[][]{comp--obs--12z}. 
The pre-maximum spectra of SN~2005hk were taken on MJD = 53679.4 \citep{2005hk--spec--pol} and  MJD = 53688.2 \citep{05hk--deflag?}, which correspond to phases of $-$12.26$\pm$0.23 and $-$3.46$\pm$0.23, respectively. Therefore, secured by the light curves, we can rule out phasing errors as the source of the difference between the two spectra. 

\par
\subsection{Existing explosion models}
As discussed previously, (§\ref{sect:intro}), several explosion scenarios have been considered for SNe Iax. Many of these are variants of explosion models first proposed for SNe Ia. These include both deflagration models \citep{nomoto-w7,read--02cx--spectra}, and delayed detonation models in which an initial deflagration transitions to a detonation \citep{khokhlov-91,hoeflich-95,ia--antideflag}. For both of these classes of models, it has been suggested that the fainter peak magnitudes and lower ejecta velocities observed in SNe Iax relative to normal SNe Ia may be explained by relatively low energy explosions. Indeed, the least energetic of these pure deflagration models succeed in only partially disrupting the white dwarf \citep{kicked--remnants,3d--deflag--sim--obs}.
\par
It has been demonstrated that both explosion scenarios naturally allow for variations in observational quantities since differences in the explosion parameters such as the ignition configuration in deflagration models \citep{3d--deflag--sim--obs}, or variations in the adopted deflagration-to-detonation transition density \citep[e.g.][]{hoeflich-02} can give rise to a range of $^{56}$Ni-masses ($\sim 0.03 - 0.6\,M_{\odot}$), which ultimately controls most of the observed properties, especially at early epochs. 
\par
However, existing models do not explore the parameter space sufficiently to ascertain whether any specific scenario is able to simultaneously account for the spectral differences observed in PS1-12bwh and SN~2005hk, as described in §\ref{sect:spectroscopic}, {\it without} significant differences in their respective light curves. For example, for SNe Ia, \cite{hoeflich-95} investigated whether variations in the initial conditions of the progenitor white dwarf could lead to observable differences for pulsational delayed detonation models. They find that it is possible to produce similar light curves with different initial conditions and explosions. However, it is unclear whether this could explain the spectroscopic observations of SN~2005hk and PS1-12bwh.
\par
In their study of deflagration models, \cite{3d--deflag--sim--obs} carried out a sequence of simulations in which the strength of the explosion was varied by altering the ignition configuration geometry. This resulted in a significant variation in $^{56}$Ni masses ranging from $\sim 0.03 - 0.3\,M_{\odot}$. They also investigated whether a relatively large change in the white dwarf central density could produce noticeably different explosions. Although the models they considered (N100Hdef\footnote{The naming scheme adopted by \cite{3d--deflag--sim--obs} is as follows: the `X' in NXdef refers to the number of sparks used to ignite the model. The peak brightness of these models typically scales with the number of ignition sparks, from $-16.8 \lesssim M_V \lesssim -19.0$. H/L refer to models with a higher or lower central density, respectively, relative to the standard N100def model.} and N100Ldef) are too bright at peak (by $\gtrsim$1 mag.) for direct comparisons with PS1-12bwh, they found that the central density does not very significantly affect the observables. Thus, for the pure deflagration models currently available, it appears unlikely that either the $^{56}$Ni mass, or variations in the central density could be the discriminating factor between SN~2005hk and PS1-12bwh. Comparable studies for fainter models are not available, but merit investigation.
\par
The multi-dimensionality of the \cite{3d--deflag--sim--obs} deflagration models  allows us to investigate whether orientation effects could explain the differences between SN~2005hk and PS1-12bwh, at least for this class of model: although the synthetic spectra computed from these models show small variations as a function of angle at all epochs, these are not sufficiently strong to significantly alter any of the spectral features discussed in §\ref{sect:spectroscopic}. For example, across the full range of viewing angles for the N5def model, the standard deviation from the average equivalent width of the \ion{Fe}{iii} $\lambda$4404 absorption profile is less than 10\%.\footnote{As the variations with viewing angle are more significant at shorter wavelengths, we focussed our attention on the prominent \ion{Fe}{iii} $\lambda$4404 absorption feature.} We therefore can not ascribe the apparent discrepancies between the pre-maximum spectra of SN~2005hk and PS1-12bwh to orientation effects in the pure deflagration models.
\par
Thus, based on existing explosion models, we cannot easily identify a parameter that drives the observed pre-maximum spectroscopic differences in SN~2005hk and PS1-12bwh. Future simulations that further explore the effects of variations in the white dwarf initial conditions would be needed to investigate this further in relation to specific explosion scenarios. In the remainder of this study, we pursue the alternative strategy of developing empirical models to better characterise the differences in ejecta properties that might be most consistent with the data.

\begin{figure*}
\centering
\includegraphics[width=\textwidth]{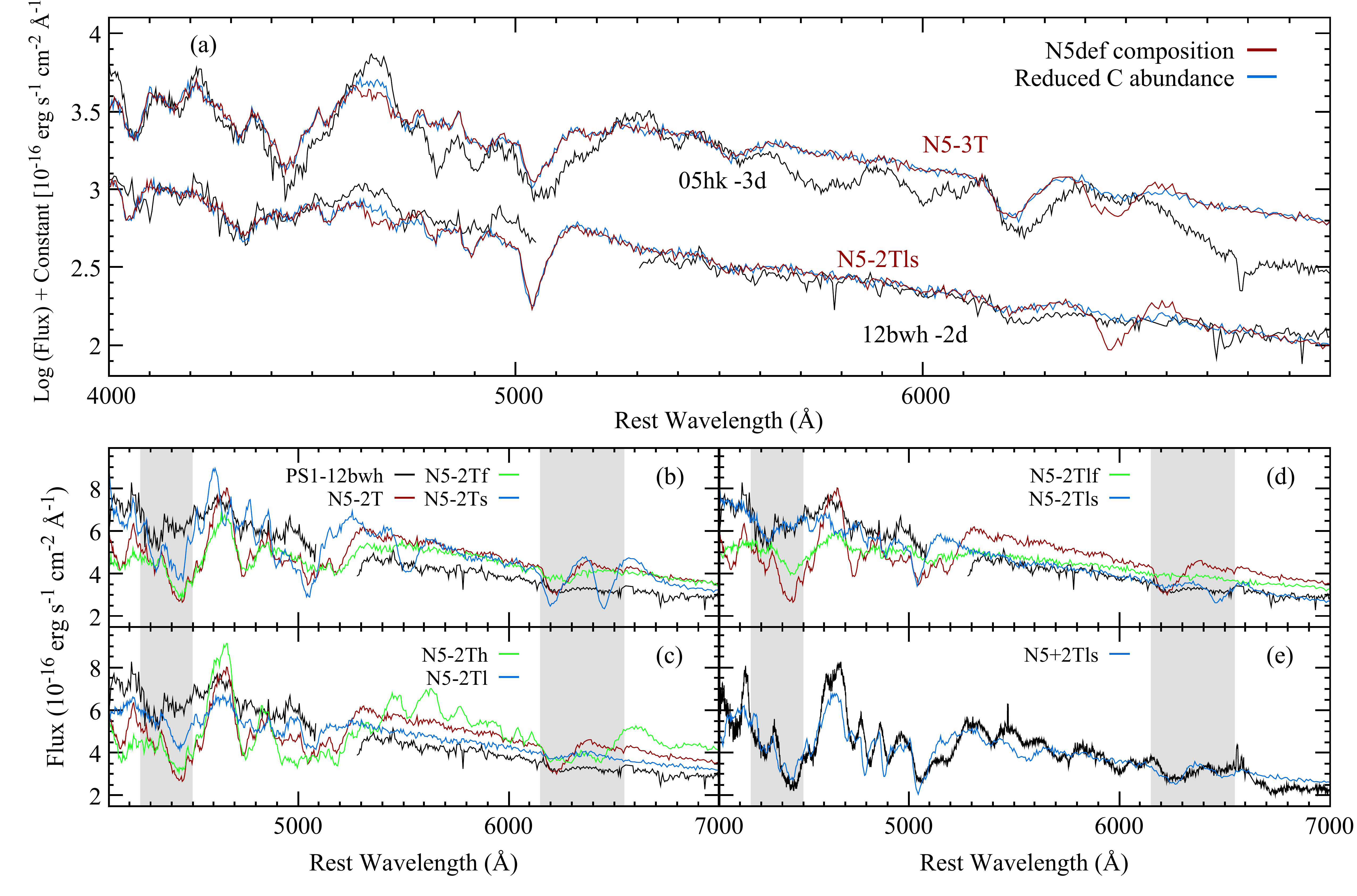}
\caption{Spectroscopic comparison of PS1-12bwh and SN~2005hk to TARDIS model spectra. The spectra of PS1-12bwh have been binned to $\Delta\lambda$ = 5, and the narrow host emission lines have been removed. Panel (a) shows the comparison of our favoured models in red (N5-3T and N5-2Tls) to observations in black. Spectra in blue show our favoured models with a carbon abundance reduced by an order of magnitude. The only difference in the resultant spectrum is the decreased strength of the \ion{C}{ii} $\lambda$6580 feature. In panels (b), (c), and (d), the N5-2T model is shown for reference. Green and blue spectra demonstrate the effects of changing certain physical parameters (see Table~\ref{tab:12bwh--models}) on the resultant synthetic spectrum. As in panel (a), comparisons are to the $-$2\,d PS1-12bwh spectrum. Shaded regions indicate the \ion{Fe}{ii} and \ion{Fe}{iii}, and \ion{Si}{ii} and \ion{C}{ii} features that we use to define a satisfactory model (see \S\ref{sect:emp-model}). Panel (e) shows a direct comparison of the N5+2Tls model spectrum to the $+2$\,d spectrum of PS1-12bwh. 
}
\label{fig:12bwh--models}
\end{figure*}

\par
\subsection{Empirical modelling}
\label{sect:emp-model}
From Fig.~\ref{fig:12bwh--residual}, it can be seen that there is a clear difference in ionisation state between the SN~2005hk $-$3\,d and the PS1-12bwh $-$2\,d spectra. For example, the \ion{Fe}{ii} $\lambda$4555 absorption is stronger than \ion{Fe}{iii} $\lambda$4404 in the SN~2005hk spectrum, while the converse is true for the PS1-12bwh spectrum. Furthermore, SN~2005hk shows strong absorption features due to \ion{Si}{ii} and \ion{C}{ii} while these are weak in the PS1-12bwh spectrum. Yet Fig.~\ref{fig:12bwh--lc} and Table~\ref{tab:12bwh--lc--param} clearly demonstrate that both objects have remarkably similar light curves. To investigate the potential parameters that may give rise to these spectroscopic differences, we used TARDIS\footnote{ http://tardis.readthedocs.org/en/latest}, a 1D Monte Carlo radiative transfer code \citep{tardis}. Our methodology is as follows: we began by constructing a model that could adequately reproduce the $-$3\,d spectrum of SN~2005hk. We use this model as an anchor against which to compare the effects of changing various physical parameters -- such as the inner ejecta boundary -- such that the resulting synthetic spectrum resembles the $-$2\,d spectrum of PS1-12bwh. We stress that this is a qualitative exercise. We further stress that, by design, the radiative transfer modelling is sensitive only to the spectral forming region, and therefore does not constrain the properties of the ejecta that lie outside this region. The details of the model parameters described in this section are listed in Table~\ref{tab:12bwh--models}. 

\begin{table*}
\centering
\caption{Input parameters for TARDIS model spectra }\tabularnewline
\label{tab:12bwh--models}\tabularnewline
\begin{tabular}{lllllll}\hline
\hline
 Model & Luminosity & Time since explosion & Inner boundary & Outer boundary & Density & Composition \tabularnewline
 & $\log L [L_{\odot}$] & $t_{\mathrm{exp}}$ (days) & $v_i$ (km s$^{-1}$) & $v_o$ (km s$^{-1}$)  & &  \tabularnewline
\hline
\hline
N5-3T	& 9.10		 	 & 19				& 7800			 & 9400			  & N5def			    & N5def \tabularnewline
N5-2T	& 8.95			 & 20				& 7800			 & 9400			  & N5def 			 	& N5def \tabularnewline
N5-2Tf	& 8.95			 & 20				& 8800			 & 9400			  & N5def 			 	& N5def \tabularnewline
N5-2Ts	& 8.95			 & 20				& 5800			 & 9400			  & N5def 			    & N5def \tabularnewline
N5-2Th	& 8.95			 & 20				& 7800			 & 9400			  & N5def $\times$ 5 	& N5def \tabularnewline
N5-2Tl	& 8.95			 & 20				& 7800			 & 9400			  & N5def $\times$ 0.2 	& N5def \tabularnewline
N5-2Tlf	& 8.95			 & 20				& 8800			 & 9400			  & N5def $\times$ 0.2 	& N5def \tabularnewline
N5-2Tls	& 8.95			 & 20				& 5800			 & 9400			  & N5def $\times$ 0.2	& N5def \tabularnewline
N5+2Tls	& 8.85			 & 24				& 5200			 & 9400			  & $v \textgreater$ 5800 km s$^{-1}$: N5def $\times$ 0.2	& N5def \tabularnewline
	& 			 & 				& 			 & 			  					  & $v \textless$ 5800 km s$^{-1}$: N5def 	&  \tabularnewline

\hline
\end{tabular}
\tablefoot{For the density and composition, we take a spherical average of velocity shells from the N5def model of \cite{3d--deflag--sim--obs}. The naming scheme of our models is defined as follows: N5-XT refers to TARDIS models (`T') which are based on the N5def model from \cite{3d--deflag--sim--obs}, with `-X' referring to the phase of the spectrum. These times  correspond to phases of $-3$\,d, $-2$\,d, and $+2$\,d relative to $r-$band maximum, assuming a rise time of $\sim$22\, \citep{05hk--deflag?, 15h}. The suffixes f/s refer to the photospheric velocity being fast/slow, while h/l refer to the ejecta density being higher/lower relative to the N5def model.
}
\end{table*}
\par

\subsubsection{Modelling the $-3$\,d SN~2005hk spectrum}
\label{05hk-model}
As input for our spectral synthesis calculations, we require both a luminosity and time since explosion for each spectrum; for SN~2005hk we determine these from the light curves of \cite{comp--obs--12z}: we set the luminosity to $L = 10^{9.1}$\,$ L_{\odot}$ and the time since explosion, $t_{\mathrm{exp}}$ = 19\,d, for our SN~2005hk model and treat both as fixed parameters. 
\par
The N5def and N3def models described by \cite{3d--deflag--sim--obs} have previously been shown to provide reasonable matches to the observables of SNe 2005hk \citep{3d--deflag--sim--rem} and 2015H \citep{15h}. Specifically, these models are able to reproduce important properties of the early spectra of both SNe, including the low line velocities and weak IME features. The corresponding model light curves show good agreement with the peak absolute magnitudes, but decline faster than observed in the redder bands. This may be the result of a lack of $\gamma$-trapping and could indicate that the model ejecta masses are too low, as discussed by \cite{3d--deflag--sim--rem}. Indeed, this model results in the production of $\sim$0.4 M$_{\odot}$ of unbound ejected material and a $\sim$1\,M$_{\odot}$ bound remnant, while \cite{obs--05hk--08a} estimate SN~2005hk produced roughly twice as much ejecta. Given the reasonable agreement between the N5def model and SN~2005hk, we opted to determine the density and composition in our model by taking averages of the N5def model in spherical velocity shells. 
\par
The only remaining free parameter for the spectral synthesis calculation is the range of velocities included in the line forming region of our model (i.e. the choice of the inner and outer boundaries of the computational domain). This range (7800 -- 9400 km s$^{-1}$) was selected to satisfactorily reproduce the observed spectrum. The resulting synthetic spectrum is shown in Fig.~\ref{fig:12bwh--models}(a), and we denote this model as N5-3T (see Table \ref{tab:12bwh--models}). 
\par
Although there appears to be a slight difference in velocity, the N5-3T model is able to broadly match many of the features observed in SN~2005hk at $-$3\,d, including the overall flux level, the strong absorption seen at $\sim$5000\,\AA, the numerous features observed at near-UV wavelengths, the ratio of \ion{Fe}{iii}\,$\lambda$4404 to \ion{Fe}{ii}\,$\lambda$4555, and the relatively strong \ion{Si}{ii} $\lambda$6355. Indeed, the profiles of \ion{Fe}{iii}\,$\lambda$4404, \ion{Fe}{ii}\,$\lambda$4555, and \ion{Si}{ii} $\lambda$6355 clearly demonstrate the differences between the spectra of SN~2005hk and PS1-12bwh. Based on these features, we deem the N5-2T model to provide reasonable agreement with SN~2005hk.
\par
The \ion{C}{ii}\,$\lambda$6580 absorption feature (Fig.~\ref{fig:12bwh--models}a) produced by N5-2T is much more pronounced in the model than in the observed spectrum. This mismatch can also be seen in the comparison of the spectrum resulting from the N5def model for SN~2005hk by \citet[][their Fig. 5]{3d--deflag--sim--rem}. This may be because the amount of carbon entrained in the N5def explosion model is too high to match SN~2005hk. We therefore reduced the carbon abundance in the ejecta of our N5-3T model by an order of magnitude, and find an improved match to the weak \ion{C}{ii} $\lambda$6580 feature, with little effect on the rest of the spectrum. Models with reduced carbon abundances are shown in blue in Fig.~\ref{fig:12bwh--models}(a). 

\par
\subsubsection{Modelling the $-2$\,d PS1-12bwh spectrum}
\label{12bwh-model}
Having generated an acceptable model for SN~2005hk at $-3$\,d, we use it as a starting point to identify the possible cause of the differences present in the line-forming regions of the two supernovae. In order to identify a model as satisfactory, we define the following basic criteria that it must meet:
\begin{itemize}
	\item a lack of lines at wavelengths long-ward of $\sim$5000\AA, with exceptions being the weak features due to \ion{Si}{ii} and \ion{C}{ii}; 
	\item the peak of the SED must occur at relatively blue wavelengths ($\sim$4000\,\AA); and
	\item deeper \ion{Fe}{iii}\,$\lambda$4404 relative to \ion{Fe}{ii}\,$\lambda$4555.
\end{itemize}
If a model satisfies these criteria, we expect that it closely approximates the ionisation state of the ejecta. Further discrepancies between the model and observations may of course arise due to differences in the exact density or composition structure, but such an investigation is beyond the scope of the broad, exploratory study here. 
\par
We begin by altering the luminosity and phasing of the N5-3T model generated for the $-$3\,d spectrum of SN~2005hk, such that they correspond to the values appropriate for PS1-12bwh at $-$2\,d. The luminosity is set to $L = 10^{8.95}$\,$L_{\odot}$, somewhat lower than SN~2005hk (0.15 dex). Given the similarity of their light curves at post-maximum epochs, we assume that their pre-maximum light curve evolution is also comparable; this is in fact supported by the limited pre-maximum coverage of PS1-12bwh and the similar post-maximum decline, and hence we set $t_{\mathrm{exp}} = 20$\,d (i.e., one day later than our N5-3T model or two days before maximum light). These two changes to the N5-3T model form our N5-2T model, which is replicated in red in panels (b), (c), and (d) of Fig.~\ref{fig:12bwh--models}. It shows strong absorption at near-UV wavelengths that is not observed in PS1-12bwh. It also produces stronger \ion{Si}{ii} absorption than is observed, and stronger \ion{Fe}{ii} compared to \ion{Fe}{iii}, which is the opposite to what is observed. Thus, the N5-3T and N5-2T models demonstrate that the differences between SN~2005hk at $-3$\,d and PS1-12bwh at $-2$\,d cannot be accounted for by simply adjusting for the slight differences in epoch and luminosity. 
\par
We tested a variety of models with shorter rise times than the 21.8\,d observed for SN~2005hk \citep{05hk--deflag?,15h}. We find that the pre-maximum spectrum of PS1-12bwh cannot be accounted for simply by shortening the rise time to peak. In addition, a significantly shorter rise time would appear to contradict the similar post-maximum evolution observed in SN~2005hk and PS1-12bwh. Furthermore we find that we are not able to produce satisfactory matches to the $+2$\,d spectrum of PS1-12bwh when assuming such a rise time. We therefore conclude that the discrepancy between pre-maximum spectra is unlikely to be due to a difference in rise times.
\par
Below we investigate changes in other physical parameters that could shed light on the pre-maximum spectra of SN~2005hk and PS1-12bwh.

\par

As we did previously for the SN~2005hk model above (N5-3T), we treat the original N5def composition as being fixed for the construction of the PS1-12bwh models. We did consider alternative compositions having up to an order of magnitude increase in IGE abundances, but these did not produce the desired results. Consequently, we retain the original N5def composition and focus our analysis on other physical parameters.

\par
We next considered the effect of changing the location of the photosphere in our spectrum calculation i.e., effectively altering the boundary radiation temperature of the model; this is shown in Fig.~\ref{fig:12bwh--models}(b). Increasing the photospheric velocity (N5-2Tf), and therefore decreasing the temperature, produces a similar spectrum to our N5-2T model but causes the \ion{Si}{ii} to become much weaker. With this increased velocity, the ratio of \ion{Fe}{iii} to \ion{Fe}{ii} remains practically unchanged. Lowering the photospheric velocity (N5-2Ts), and therefore increasing the temperature increases the strength of the \ion{C}{ii}, \ion{Si}{ii}, and \ion{S}{ii}. Therefore, the placement of the photosphere in the model alone cannot easily account for the observed differences.

\par
At early epochs, it is the outermost layers of the SN ejecta that influence the appearance of the observed spectra. As the differences between SN~2005hk and PS1-12bwh are limited to pre-maximum epochs, it is likely that the main differences also occur in the outermost ejecta layers. We investigate this by altering the density in the outer regions. We constructed density profiles within the region delimited by 5800 -- 9400 km~s$^{-1}$ that are a factor of five higher or lower than that of the N5def model. We stress that this is the region of ejecta probed by our models, and we do not comment on ejecta properties outside this region. For the models considered in this study, these regions correspond to $\sim25 - 35$\% of the ejecta mass. We stress that we have taken the relatively simple approach of scaling the total density in the spectrum-forming region by an overall factor.  Although simplistic, this approach allows us to easily explore the sensitivity to density; given the limited quality and time-coverage of our spectral series, more detailed attempts to pin down an exact density profile are not warranted here.
\par

At higher densities (N5-2Th) relative to N5def, the model produces many strong absorption features between $\sim5000 - 6000$\,{\AA}, violating our first criterion for an acceptable match above. The lower density model (N5-2Tl) however, matches the redder wavelengths, with only weak \ion{Si}{ii} and \ion{C}{ii} absorption present. N5-2Tl also matches the \ion{Fe}{iii} $\lambda$4404 profile, but produces stronger \ion{Fe}{ii} $\lambda$4555 absorption. We therefore conclude that while a lower density for high velocity ejecta improves upon the N5-2T model, it alone is not sufficient to match the observed spectrum of PS1-12bwh.

\par
Given the rough agreement with the data afforded by a lower density (N5-2Tl), we now test whether changes in photospheric velocity, in conjunction with a lower density, might further improve upon the N5-2Tl model. Doing so yields synthetic spectra that are shown in Fig.~\ref{fig:12bwh--models}(d). As mentioned previously, an increased photospheric velocity (N5-2Tf) causes a desirable reduction in the strengths of features at longer wavelengths - this is also observed in the N5-2Tl model. However, simultaneously implementing a lower density and increasing the photospheric velocity (N5-2Tlf) is not able to overcome the shortcomings of either parameter on its own i.e., the N5-2Tlf model does not reproduce the \ion{Fe}{iii} $\lambda$4404 to \ion{Fe}{ii} $\lambda$4555 ratio in the correct sense. Instead, we find that the model that is best able to meet all three criteria described above is one that simultaneously incorporates a lower density for high velocity ejecta and a lower photospheric velocity (N5-2Tls). Indeed, a lower photospheric velocity may be seen as a natural consequence of a reduced density, as the photosphere recedes faster through the less dense ejecta. Therefore these parameters are qualitatively consistent with each other.

\par
We are now in a position to conclude that the divergence between the $-3$\,d spectrum of SN~2005hk and $-2$\,d spectrum of PS1-12bwh can be attributed to differences in their ejecta structures. Specifically, differing densities of the high velocity material (by approximately a factor of a few). Such a scenario naturally explains why this only manifests in the earliest spectra, as it is at these epochs that the highest velocity ejecta are observed. We speculate that the density of the lower velocity material may be comparable for both objects, hence the spectroscopic similarity at later epochs. 

\par

In order to test this, we modelled the $+2$\,d spectrum of PS1-12bwh and allowed for material at lower velocities to have a higher density. We adjusted the time since explosion ($t_\mathrm{exp}$ = 24\,d), and slightly decreased the luminosity ($L = 10^{8.85} L_{\odot}$) and inner velocity boundary ($v_{\mathrm{i}} = 5200$\,km~s$^{-1}$). For the ejecta above 5800\,km~s$^{-1}$, we maintained the same density profile as our N5-2Tls model, but used an unmodified N5def density profile between 5800 and 5200\,km~s$^{-1}$; the resulting synthetic spectrum compared to PS1-12bwh is shown in Fig.~\ref{fig:12bwh--models}(e). Thus, with these adjustments, we were able to reproduce the correct ratio of \ion{Fe}{iii} $\lambda$4404 to \ion{Fe}{ii} $\lambda$4555 at $+2$\,d (which has inverted since $-2$\,d) and generate more prominent features at redder wavelengths than are visible at $-2$\,d. Therefore, accepting that the N5+2Tls model results in a spectrum that is broadly consistent with the  $+2$\,d spectrum of PS1-12bwh, we can conclude that this spectrum is consistent with the higher velocity ejecta of PS1-12bwh having a lower density compared to SN~2005hk.

\par
It is now clear that there are additional factors that drive the diversity of SNe Iax. Typically, variations in thermonuclear explosions are attributed to difference in total ejecta mass or $^{56}$Ni yield. The similarity in light curve properties of PS1-12bwh and SN~2005hk, however, implies that these bulk properties are similar for the two supernovae. Therefore, the difference in pre-maximum spectra is likely due to an additional factor that is not highly correlated with the amount of $^{56}$Ni produced. We speculate that differences in the initial conditions of the progenitor white dwarf or ignition properties (e.g. location and configuration of thermonuclear runaway) can lead to different ejecta structures while producing comparable amounts of $^{56}$Ni. Further explosion model simulations are required in order to test whether such properties produce variations similar to those observed in SN~2005hk and PS1-12bwh.

\par

%

\section{Summary}
\label{sect:sum}
In this study we presented photometric and spectroscopic observations of PS1-12bwh, a Type Iax supernova. We find its light curve to be almost identical to that of SN~2005hk, with only small differences in peak luminosities and decline rates. 
\par
At later epochs ($\gtrsim$1 month post-maximum light) PS1-12bwh shows spectroscopic similarities to SN~2005hk, as well as other SNe Iax. Our earliest spectrum of PS1-12bwh, however, appears quite dissimilar to spectra of SN~2005hk at comparable epochs and bears a much closer resemblance to spectra with phases approximately one week earlier. We investigated possible factors that may explain the difference in ionisation state between the two similarly phased spectra of SN~2005hk and PS1-12bwh. We find a likely explanation to be that both objects have different amounts of high velocity ejecta, with PS1-12bwh also having a lower photospheric velocity (by $\sim$2000 km s$^{-1}$) at $-$2\,d than SN~2005hk at $-$3\,d. This would naturally explain why the differences are apparent only in the pre-maximum spectra, with the post-maximum spectra being similar both to each other, as well as other SNe Iax with comparable lightcurve properties. 
\par
SN~2005hk and PS1-12bwh further underline the heterogeneous nature of SNe Iax. As a corollary, our study highlights the difficulty in assigning spectroscopically estimated epochs to SNe Iax. Without light curve information, the phases inferred from spectroscopic comparisons may be mis-estimated by a week or possibly more, thereby potentially affecting the classification itself, or follow-up of SNe Iax targets. Our study highlights the need for further follow-up of type Iax SNe on the one hand, and investigation into different initial conditions of the progenitor white dwarf on the other.

%

\begin{acknowledgements}
We thank the anonymous referee whose comments motivated us to revisit and clarify aspects of our manuscript. We also thank the authors of \cite{3d--deflag--sim--obs} for making the N5def explosion model available to us.
RK and SAS acknowledge support from STFC via ST/L000709/1. 
The UCSC group is supported in part by NSF grant AST-1518052 and from fellowships from the Alfred P.\ Sloan Foundation and the David and
Lucile Packard Foundation to RJF. Some of the observations reported here were obtained at the MMT Observatory, a joint facility of the Smithsonian Institution and the University of Arizona. The Pan-STARRS1 Surveys (PS1) have been made possible through contributions of the Institute for Astronomy, the University of Hawaii, the Pan-STARRS Project Office, the Max-Planck Society and its participating institutes, the Max Planck Institute for Astronomy, Heidelberg and the Max Planck Institute for Extraterrestrial Physics, Garching, The Johns Hopkins University, Durham University, the University of Edinburgh, Queen's University Belfast, the Harvard-Smithsonian Center for Astrophysics, the Las Cumbres Observatory Global Telescope Network Incorporated, the National Central University of Taiwan, the Space Telescope Science Institute, the National Aeronautics and Space Administration under Grant No. NNX08AR22G issued through the Planetary Science Division of the NASA Science Mission Directorate, the National Science Foundation under Grant No. AST-1238877, the University of Maryland, and E\"{o}tv\"{o}s Lor\'{a}nd University (ELTE) and the Los Alamos National Laboratory. The Liverpool Telescope is operated on the island of La Palma by Liverpool John Moores University in the Spanish Observatorio del Roque de los Muchachos of the Instituto de Astrofisica de Canarias with financial support from the UK Science and Technology Facilities Council (proposal ID: PL12B05, PL15A22; P.I.: R. Kotak). Images taken with Faulkes Telescope North operated by Las Cumbres Observatory Global Telescope Network (proposal ID: LCO2012B-010; P.I.: S. J. Smartt). 
\end{acknowledgements}

\bibliographystyle{aa}
\bibliography{Magee_SNeIax-12bwh}

\begin{appendix}
\onecolumn
\section{Tables}
\label{apdx:tables}

\begin{table}[!ht]
\centering
\caption{Local sequence stars used to calibrate PS1-12bwh photometry.}\tabularnewline
\label{tab:12bwh---std}\tabularnewline
\begin{tabular}{lllllll}\hline
\hline
\tabularnewline[-0.25cm]
No.  & RA & DEC &  $g$ & $r$ & $i$ & $z$  \tabularnewline
& &  &  (mag) & (mag) & (mag) & (mag) \tabularnewline
\hline
\hline
\tabularnewline[-0.25cm]
\phn1 & 07:09:24.46 & +39:05:49.4 & 17.52(0.01) & 17.09(0.01) & 16.94(0.01) &  16.88(0.01)  \tabularnewline
\phn2 & 07:09:31.70 & +39:05:41.3 & 18.97(0.01) & 18.49(0.01) & 18.31(0.01) &  18.20(0.03)  \tabularnewline
\phn3 & 07:09:31.64 & +39:06:34.7 & 18.46(0.01) & 17.97(0.01) & 17.80(0.01) &  17.73(0.02)  \tabularnewline
\phn4 & 07:09:30.85 & +39:06:52.1 & 17.63(0.01) & 17.24(0.01) & 17.11(0.01) &  17.08(0.01)  \tabularnewline
\phn5 & 07:09:28.60 & +39:06:53.5 & 16.38(0.01) & 15.83(0.01) & 15.65(0.01) &  15.56(0.01)  \tabularnewline
\phn6 & 07:09:25.13 & +39:07:41.3 & 17.17(0.01) & 16.74(0.01) & 16.59(0.01) &  16.50(0.01)  \tabularnewline
\phn7 & 07:09:19.83 & +39:08:21.8 & 18.06(0.01) & 17.61(0.01) & 17.45(0.01) &  17.38(0.02)  \tabularnewline
\phn8 & 07:09:11.96 & +39:04:45.8 & 18.30(0.01) & 17.63(0.01) & 17.39(0.01) &  17.27(0.01)  \tabularnewline
\phn9 & 07:09:15.13 & +39:04:27.0 & 17.62(0.01) & 17.04(0.01) & 16.79(0.01) &  16.66(0.01)  \tabularnewline
10 	  & 07:09:18.76 & +39:04:09.0 & 17.79(0.01) & 17.21(0.01) & 17.02(0.01) &  16.90(0.01)  \tabularnewline
\hline
\end{tabular}
\tablefoot{Magnitudes of sequence stars are taken from SDSS-DR9 and shown to two decimal places in the AB system. 1$\sigma$ uncertainties are given in parentheses.}
\end{table}

\begin{table}[!h]
\centering
\caption{Spectroscopy log for PS1-12bwh}\tabularnewline
\label{tab:12bwh--specseq}\tabularnewline
\begin{tabular}{clllll}\hline
\hline
\tabularnewline[-0.25cm]
Date  & MJD & Phase &  Telescope +  & Grating & Wavelength coverage 
\tabularnewline
& & (days) &  Instrument &  & \hspace{1.5cm}($\AA$)  \tabularnewline
\hline
\hline
\tabularnewline[-0.2cm]
2012 Oct. 22 & 56223.14 & \phn$-$2   & WHT+ISIS  	& R300B \& R158R 	& 3500 - 5200, 5400 - 9300   \tabularnewline 
2012 Oct. 26 &     56226.64  & \phn+2	& Gemini + GMOS-N & B600/450 \& R400/750 & 3600 - 9700 \tabularnewline
2012 Nov. 14 & 56245.40 & $+21$ 	& MMT$+$BlueChannel & 300GPM & 3300$-$8500 \tabularnewline
2012 Nov. 20 & 56252.13 & +27		& WHT+ISIS	 	& R300B \& R158R 	& 3500 - 5200, 5400 - 9300   \tabularnewline
2012 Dec. 20 & 56282.09 & +57		& WHT+ISIS 		& R300B \& R158R	& 3500 - 5200, 5400 - 9300   \tabularnewline

\hline
\end{tabular}
\end{table}

\begin{table}[!h]
\centering
\caption{References for comparison SNe spectra used throughout this paper.}\tabularnewline
\label{tab:objs}\tabularnewline
\begin{tabular}{llllll}\hline
\hline
\tabularnewline[-0.25cm]
SN  & SN Type & M$_r$ & $\Delta$m$_{15}$(r) &  E(B$-$V)$_{host}$ & Reference \tabularnewline
\hline
\hline
\tabularnewline[-0.25cm]
2005hk		& Iax 	& $-18.07\pm$0.25 & 0.70$\pm$0.02 & 0.11 &	1, 2, 3, 4	\tabularnewline
2008ha		& Iax	& $-15.15\pm$0.14 & 1.11$\pm$0.04 & $-$ &   5	\tabularnewline
2012Z		& Iax	& $-18.60\pm$0.09 & 0.66$\pm$0.02 & 0.11 &	4	\tabularnewline
2015H		& Iax	& $-17.27\pm$0.07 & 0.69$\pm$0.04 & $-$ &	6	\tabularnewline
\hline
\end{tabular}
\tablefoot{(1) \cite{05hk--deflag?}; (2) \cite{2005hk--spec--pol}; (3) \cite{05hk--blondin}; (4) \cite{comp--obs--12z}; (5) \cite{obs--10ae} (6) \cite{15h}. All spectra were obtained from WISeREP \citep[][, http://wiserep.weizmann.ac.il]{wiserep}.}

\end{table}

\end{appendix}

\end{document}